\documentclass[preprint]{aastex}
\shortauthors{Xiang et al.}
\usepackage{multirow}
\usepackage{epsfig}
\usepackage{subfigure}
\usepackage{graphicx,times}
\input{epsf.sty}                        

\begin{document}
\title{The Centennial Evolution of Solar Chromospheric Rotation}
\author{N. B. Xiang\altaffilmark{1,2}, X. H. Zhao\altaffilmark{2}, L. H. Deng\altaffilmark{3}, F. Y. Li\altaffilmark{4,2}, M. Wan\altaffilmark{3}, S. Y. Qi\altaffilmark{5,2}}
\affil{$^{1}$Yunnan Observatories, Chinese Academy of Sciences, Kunming 650216, China}
 \affil{$^{2}$State Key Laboratory of Solar Activity and Space Weather, National Space Science Center, Chinese Academy of Sciences, Beijing 100190, China; xhzhao@spaceweather.ac.cn}
\affil{$^{3}$School of Mathematics and Computer Science, Yunnan Minzu University, Kunming 650504, China}
 \affil{$^{4}$Institute of Optics and Electronics, Chinese Academy of Sciences, Chengdu 610209, China}
 \affil{$^{5}$College of Science, Henan Agricultural University, Zhengzhou, Henan,450002, China}


\begin{abstract}
Rotation is a prominent feature of the Sun, and it plays a crucial role in the generation and dynamic evolution of solar magnetic fields. The daily composite time series of Ca II K plage areas from 1907 February 1 to 2023 December 31 is used to analyze its periodicity and examine the temporal variation of its rotation period lengths (RPLs) using continuous wavelet transform. Wavelet analysis reveals that over a time span of more than a century, chromospheric rotation exhibits a dominant synodic period of approximately 26.62 days, with complex temporal variations.  The long-term trend of chromospheric rotation is well-characterized by a statistically significant quadratic polynomial, showing a gradual deceleration from solar cycles 15 to 19, followed by a gradual acceleration from cycles 19 to 24. The RPLs exhibit a negative correlation between the rotation rate of the chromosphere and solar magnetic activity.  Their behavior follows a distinct pattern within a Schwabe cycle: the rotation period progressively lengthens during the initial approximately 3 years, then maintains a relatively long value from year 3 to approximately 7.5, and finally shortens during the declining phase, returning to a minimum near the subsequent solar minimum. The  variations of chromospheric RPLs show significant periods of 3.2, 5.7, 7.7, 10.3, and 12.3 years, with cross-correlation analysis pointing to a complex relationship with solar activity. The possible mechanisms for the temporal variation of the chromospheric rotation are discussed.
\end{abstract}


\section{Introduction}
The phenomenon of solar differential rotation was first discovered in 1630 by Christoph Scheiner, who noted, following Galileo's sunspot observations, that equatorial sunspots rotate faster than those at higher latitudes. Since then, solar rotation has been one of the most fundamental and enduring topics in solar physics (Howe 2009; Patern{\`o} 2010). The fundamental reason is that the interaction between the Sun's rotation and the convection zone plays a critical role in generating and dynamically evolving the solar magnetic field. This process, which is closely connected to the solar dynamo theory, in turn drives all solar activities (Babcock 1961; Thompson et al. 2003; Javaraiah et al. 2005,  Badalyan \& Obridko 2017; Cameron et al. 2017; Javaraiah 2020; Hotta \& Kusano 2021). Thus, research on solar rotation is crucial to understanding the physical mechanisms that underlie various solar activities and cycles.

Over the past century, advancements in observational technology and data accumulation have enabled extensive research on solar differential rotation across all solar layers, from the interior to the corona. Given that these layers possess distinct physical properties, studying their rotation necessitates significantly different methodologies. The rotation in the solar interior is typically determined using well-developed helioseismology techniques, such as p-mode splitting (Brown 1985; Howe 2009). For the solar atmosphere, it is widely acknowledged that magnetic feature tracking, spectroscopic measurements, and flux modulation are the three primary methods for studying the rotation. Magnetic feature tracking determines the solar rotation rate by measuring the displacement of tracers, such as sunspots, active regions, or ephemeral regions in the photosphere (Gilman 1974; Ru{\v{z}}djak et al. 2017; Kutsenko 2021; Kutsenko et al. 2023; L{\"o}{\ss}nitz et al. 2025), filaments in the chromosphere (Braj{\v{s}}a et al. 1991; Wan \& Li 2022), and coronal bright points in the corona (Braj{\v{s}}a et al. 2004; Karachik et al. 2006), over time. Spectroscopic measurements investigate solar rotation by analyzing the Doppler shift of spectral lines (Howard \& Harvey 1970; Snodgrass \& Ulrich 1990; Rao et al. 2024). The flux modulation method studies the solar rotation by applying appropriate mathematical techniques to extract the rotational signals from the flux of specific spectral lines, such as the 30.4 nm spectral line (Sharma et al. 2021; Wu et al. 2023), He I line (Li et al. 2020; Li \& Xu 2024), and Ca II K line (Wan \& Gao 2022; Li et al. 2023). These studies have consistently concluded that the upper atmosphere of the Sun, including the chromosphere, transition region, and corona, rotates faster than the underlying photosphere but exhibits a weaker degree of differential rotation.

All methods for studying solar rotation have their own limitations and errors. It is well-known that the tracer method is susceptible to the evolution and deformation of feature structures. For instance, when using sunspots as tracers to measure solar rotation, a major limitation is the scarcity of sunspots beyond a latitude of approximately $40^{\circ}$, making it difficult to obtain a reliable fit in these higher-latitude regions. Furthermore, because sunspots are anchored in deeper subphotospheric layers, they exhibit systematically higher rotation rates than the solar surface, leading to a rotation law that is measurably faster (L{\"o}{\ss}nitz et al. 2025). The Doppler method can achieve measurements at higher latitudes, but it suffers from reduced sensitivity at the disk center (Snodgrass 1984; Rao et al. 2024). The flux modulation method requires observations to satisfy two key criteria: first, the emission must originate from discrete regions of the Sun rather than being uniformly distributed across the entire solar disk; second, the radiation should not be significantly corrupted by irregular scattering during its propagation through the intervening medium. The rotation period derived through this approach represents an averaged value specific to the emitting regions on the Sun (Singh et al. 2021a). Therefore, researchers must be aware that the results obtained through different methods will differ and have a clear understanding of the specific physical quantities they aim to measure, track, or model. On the other hand, many studies have also adopted a global perspective method, utilizing various solar activity indices or the full-disk flux of spectral lines to investigate the rotation periods of different solar layers and their temporal evolution (Vats et al. 2001; Heristchi \& Mouradian 2009; Chandra \& Vats 2011; Li et al. 2019; Singh et al. 2021b; Xiang et al. 2023, 2024). This approach aims to reveal the physical mechanisms behind variations in solar rotation, thereby enhancing our understanding of solar dynamo theory. In the global perspective method, fine-scale structures, local distributions, and short-term evolutionary features are averaged out. As a result, this method remains unaffected by such factors, reflects the rate of variation in the solar atmosphere, and operates independently of the quiet Sun’s rotation (Xu et al. 2020). However, the global perspective method also has its limitations, with the most significant drawback being its inability to analyze latitude-dependent variations in solar rotation.

To date, employing a global perspective, extensive research has been carried out on the rotation of the solar atmosphere and its temporal evolution. For the photosphere, studies based on sunspots and sunspot areas have revealed that there was an accelerating trend in its rotation from the mid-19th to the early 21st century. The changes in rotation periods exhibited certain quasi-periodic variations, but showed no solar-cycle-related variability (Li et al. 2011a, b). Analysis of daily solar spectral irradiance data has demonstrated that the solar corona rotates faster than the underlying photosphere. This acceleration is primarily driven by small-scale magnetic fields, which partially reveals their role in coronal heating (Li et al. 2019; Xiang et al. 2023). In the solar transition region, Singh et al. (2021a) investigated radio flux data at multiple frequencies during the period from 1967 to 2010 and reported that the temporal evolution of the rotation period in the transition region exhibits 11-year and 22-year periodic components. Furthermore, studies using other spectral lines, such as Ly $\alpha$ emission at 121.56 nm and solar irradiance at 93.5 nm, have also revealed that the long-term variations in the rotation of the transition region exhibit an 11-year periodicity (Zhang et al. 2023; Kumar et al. 2024). In the corona, Li et al. (2012) analyzed a long-term series of 10.7-cm solar radio flux from 1947 to 2009 and identified a slight secular decrease in the coronal rotation period, though without a statistically significant 11-year Schwabe cycle. Similar declining trends were reported by Xie et al. (2017) and Deng et al. (2020), who used 10.7-cm flux and a modified coronal index spanning 1939-2019, respectively. Both studies also detected significant periodicities between 3 and 11 years in the rotation period and suggested a connection to the Schwabe cycle. This correlation, however, conflict with those of Chandra \& Vats (2011) and Li et al. (2012). More recently, Edwards et al. (2022) used tomographic maps from 2007-2020 and revealed a sharp drop in the equatorial rotation rate of the corona around 2009, followed by a recovery to faster rotation by 2017. However, based  on the global perspective, studies on the long-term evolution of chromospheric rotation remain scarce to date. The primary reason is that disk-integrated, also known as ``sun-as-a-star'', measurements in the Ca II K line only began much later, in the 1960s (Chatzistergos et al. 2020, 2024). Moreover, prior to 2022, observational data from various sources had not been cross-calibrated and combined into a continuous long-term time series. Nevertheless, studying chromospheric rotation from a global perspective is highly reliable and scientifically significant, especially since the chromosphere exhibits more rigid rotation compared to the photosphere (Li et al. 2020; Wan \& Gao 2022). To our knowledge, only Xu et al. (2020) investigated variations in chromospheric rotation using a composite Mg II index time series spanning from  1978 November to  2020 January. Their study revealed a gradual accelerating trend in chromospheric rotation over the last four solar cycles. However, the evolutionary characteristics of chromospheric rotation at other timescales, particularly its long-term variations and relationship with solar cycles, remain incompletely understood. Therefore, investigating the temporal evolution of chromospheric rotation over longer timescales, specifically at centennial timescales, is highly necessary and significant.

Building upon the consistent automated procedure established by Chatzistergos et al. (2020), Chatzistergos et al. (2024) analyzed 45 Ca II K archives to update the composite
series of the plage areas originally synthesized from corrected and calibrated historical and modern Ca II K images. This updated time series now spans from 1892 to 2023, covering the ascending phase of solar cycle 25 and representing the longest such record so far. Moreover, the method  developed by Chatzistergos et al. (2019, 2020) is the only one that has produced the plage area time series which has undergone photometric calibration. In this study, we employ the longest available composite time series of the plage areas to analyze the temporal variations of chromospheric rotation across multiple timescales and further explore the physical mechanisms underlying these variations.

\section{Temporal Evolution of the Chromospheric Rotation Period}
\subsection{Data}
The Ca II K line observations of the Sun have been conducted in two primary forms: full-disk images and disk-integrated time series. The Ca II K line was first observed using a spectroheliograph, and systematic full-disk solar photography began in 1892 (Hale, 1892; Chatzistergos et al. 2024). Subsequently, multiple observatories have conducted observations in the Ca II K line. To date, more than 40 sites worldwide have regularly performed such observations during different periods (Chatzistergos et al., 2022). Among them, some of the most prominent archives are derived from observatories such as Kodaikanal, Mt Wilson, and Coimbra (Chatzistergos et al., 2022, 2024). In contrast, disk-integrated (“sun-as-a-star”) measurements in this spectral line began considerably later, during the 1960s. Chatzistergos et al. (2020) applied a consistent automatic processing procedure, including photometric calibration (where needed), limb-darkening compensation, and artefact correction— to analyze the full-disk Ca II K observations from 38 datasets. This process enabled the production of what was at the time the most complete time series of the plage areas, covering the period from 1892 to 2019. Recently, Chatzistergos et al. (2024) reapplied this method by incorporating 7 additional full-disk Ca II K observation datasets, resulting in a total of 45 datasets. Through collective analysis, they updated the composite series of the plage areas. The newly synthesized time series now spans the period from 1892 to 2023, including the ascending phase of solar cycle 25. This new dataset has been publicly released online at https://www2.mps.mpg.de/projects/sun-climate/data.html. In this study, we utilize this updated composite time series of the plage areas to investigate the temporal evolution of the chromospheric rotation period.

\begin{figure*}[h!]
\begin{center}
\includegraphics[width=1.0\textwidth]{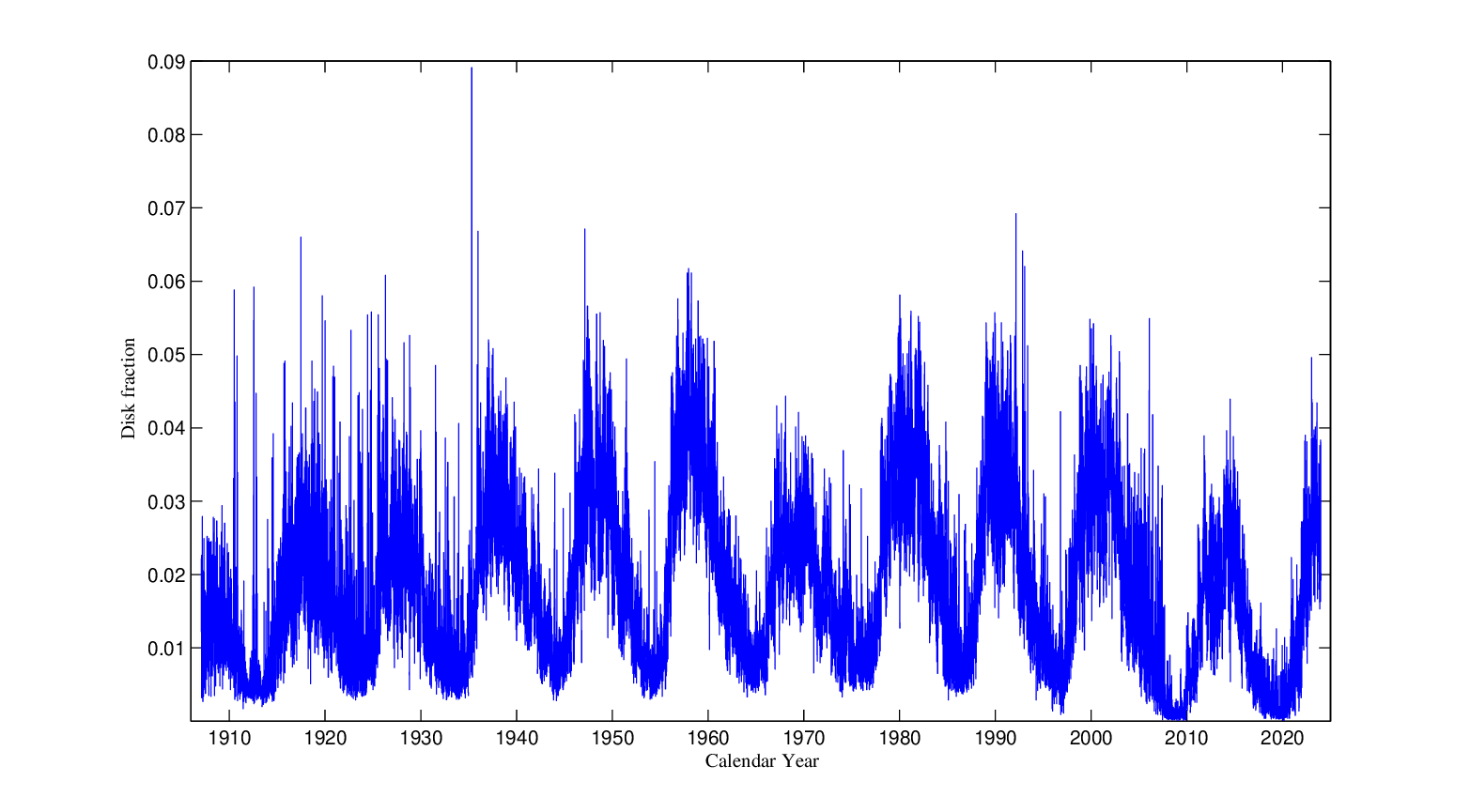}
\end{center}
\caption{The composite time series of the plage areas (in disk fraction units) from 1907 February 1 to 2023 December 31. }
\end{figure*}

The composite time series of the plage areas spans 131 years, but the entire time series has data gaps accounting for 11.59\%. However, if the data before 1907 February are removed from the entire time series, the remaining segment still spans 42,703 days, with 638 days of missing data, accounting for only 1.49\%, making it highly suitable for this study. Therefore, we utilize the composite time series of the plage areas from 1907 February 1 to 2023 December 31, as shown in Figure 1. It clearly shows that the plage area exhibits a distinct approximately 11-year solar activity cycle, with higher values and enhanced fluctuations during solar maximum. This pattern arises due to the close correlation between chromospheric plages and photospheric magnetic activity.

\subsection{Rotation Period of the Chromosphere}
In research methods for studying solar rotation, the flux modulation method primarily employs auto-correlation and wavelet transform techniques to investigate the solar rotation of various solar activity indicators (Vats et al. 2001; Chandra \& Vats 2011; Zhang et al. 2023; Xiang et al. 2024). Wavelet analysis examines power spectra to identify dominant periodicities across multiple temporal scales and localized oscillatory patterns within one-dimensional time series. This method decomposes the signal into a two-dimensional time-frequency representation, producing local wavelet power spectra that allow for precise characterization of temporal and frequency features. As a result, specific frequency components at discrete time points can be accurately identified within the analyzed series (Torrence \& Compo 1998; Grinsted et al. 2004). Thus, wavelet analysis is particularly well-suited not only for investigating the rotation period of the composite plage area time series, but also for examining the temporal evolution of its rotation characteristics.
In this study, we employ the continuous wavelet transform (CWT) using a Morlet wavelet as the mother wavelet, chosen for its effectiveness in feature extraction and optimal time-frequency localization properties (Torrence \& Compo 1998; Chowdhury \& Dwivedi 2011; Xie et al. 2017). We systematically tested parameters and found that a dimensionless frequency of $\omega_{0}=12$ offers the best balance between time and frequency resolution. The statistical significance of the wavelet power spectra was evaluated under a red-noise assumption, with periods identified using a stringent 99\% confidence level to ensure robustness.

The data gaps in the composite time series of the plage areas during the time interval we considered were filled using linear interpolation. The 638 missing data points are randomly distributed, with the longest gaps generally not exceeding 3 days, significantly shorter than the 27-day solar rotation timescale. Thus, the data gaps have minimal impact on the analysis of the rotation period after interpolation. Subsequently, we employed the CWT to analyze periodicities in the composite plage area time series, with the resulting power spectra arranged in chronological order from top to bottom panel in Figure 2. In each panel, the 99\% confidence level is indicated by black contours. As the  purpose of this study is to investigate the rotation period and its temporal evolution,  the result of wavelet analysis displays the continuous wavelet power spectra only within the timescales of 16 to 45 days.  Using the local continuous wavelet power spectra, we calculate the time-averaged wavelet spectrum over all time intervals considered. This yields the global wavelet spectra, which aid in identifying the dominant periods. The global wavelet spectrum of the composite plage area, along with its corresponding 99\% confidence level, is shown in the right column of the bottom panel of Figure 2. As shown in this figure, the continuous wavelet power spectra clearly exhibit localized oscillatory features at solar rotation timescales. This provides definitive evidence for temporal variation in the chromospheric rotation period during the interval studied. The global wavelet power spectrum exhibits narrow, steep peak, demonstrating its capability to precisely determine the dominant synodic rotation period, which is 26.62 days for the chromosphere.

\begin{figure*}[h!]
\begin{center}
\includegraphics[width=1.0\textwidth]{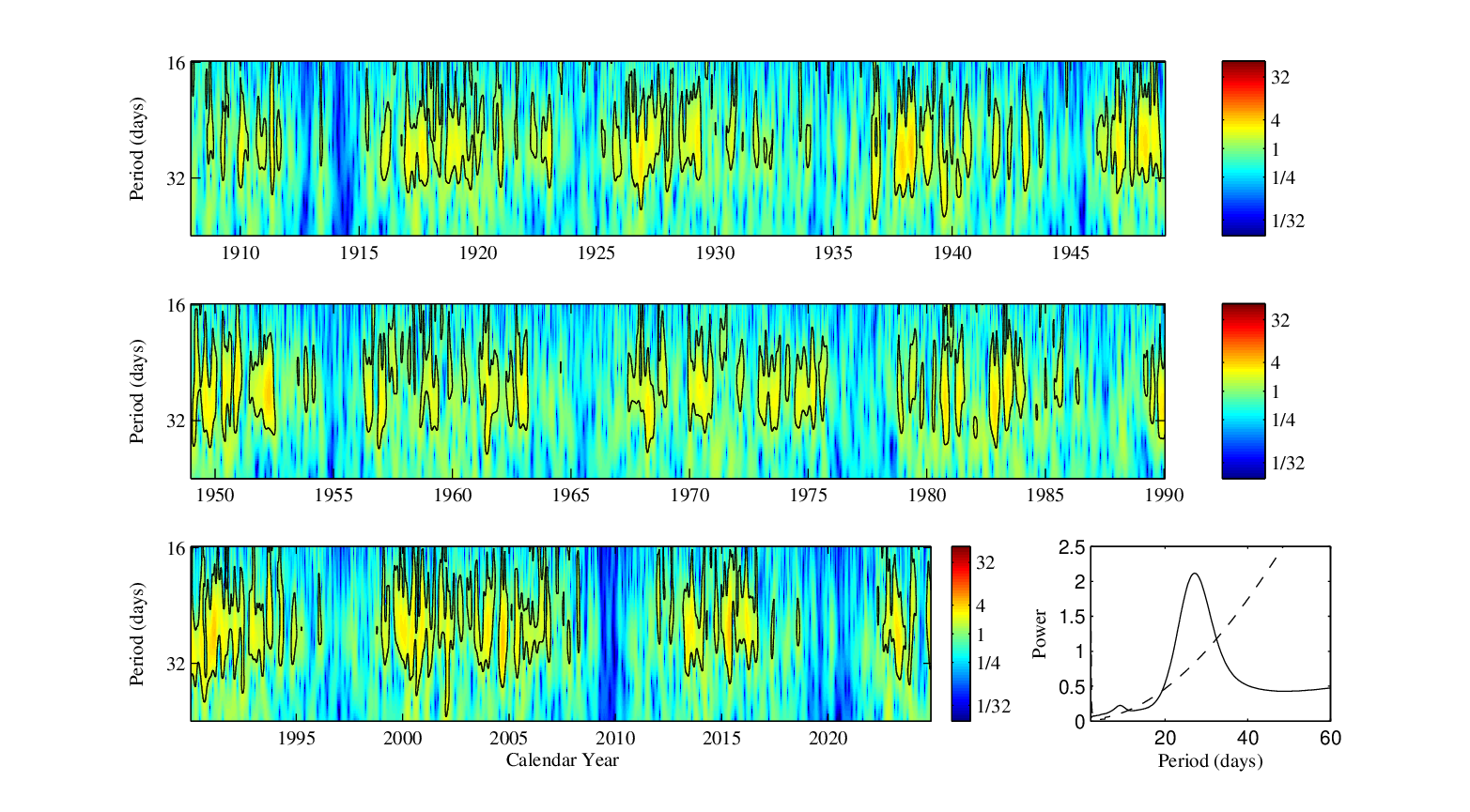}
\end{center}
\caption{Continuous wavelet power spectra of the composite time series of the plage areas from 1907 February 1 to 2023 December 31,  arranged in chronological order from top to bottom panel. In each panel, the 99\% confidence level is indicated by black contours. To clearly exhibit the rotation period and its temporal evolution, only the power spectra on the timescales of 16 to 45 days are shown. Right column of the bottom panel: global wavelet power spectrum (solid line) and the corresponding 99\% confidence level (dashed line) for the composite time series of the plage areas during the  entire  time interval considered.}
\end{figure*}

\subsection{Temporal Variation of the Rotation Period Lengths of the Chromosphere}
In order to further investigate the temporal evolution of chromospheric rotation, we conducted a detailed analysis of the local power spectrum of the plage areas shown in Figure 2, purposing to examine the local oscillatory characteristics of its rotation period. As shown in this figure, the local wavelet power spectra exhibit distinct peaks at the rotational timescales for each day. The frequencies of these peaks can be used to determine the rotation period of the plage area for each day over the entire time interval considered. The same method is widely employed in previous studies,  but the chosen timescale ranges for the rotation period vary. For instance, Li et al. (2011a) used a range of 25-31 days for sunspot areas, which was also adopted by Xie et al. (2017) for studying the temporal evolution of coronal rotation. In contrast, Katsavrias et al. (2012) reported a range of 22-30 days for the interplanetary magnetic field. Studying the adjacent transition region, Zhang et al. (2023) found a rotation period range of 22.24-31.49 days. Therefore, by integrating the results from Zhang et al. (2023) with the local characteristics of the wavelet power spectrum shown in Figure 2, we select the rotation period range of 22-32 days for the solar chromosphere. The derived daily rotation periods  are smoothed using a one-year sliding average to suppress short-term, intense fluctuations caused by strong magnetic activity. The resulting time series, shown by the solid blue line in bottom panel of Figure 3, represents the variation in the rotation period lengths (RPLs) of the plage area with solar activity. As this figure shows, the chromospheric RPLs vary on the timescales of longer than 1 yr to the solar cycle,  and that these variations appear to be quasi-periodic.

\begin{figure*}[!htbp]
\begin{center}
\includegraphics[width=0.8\textwidth]{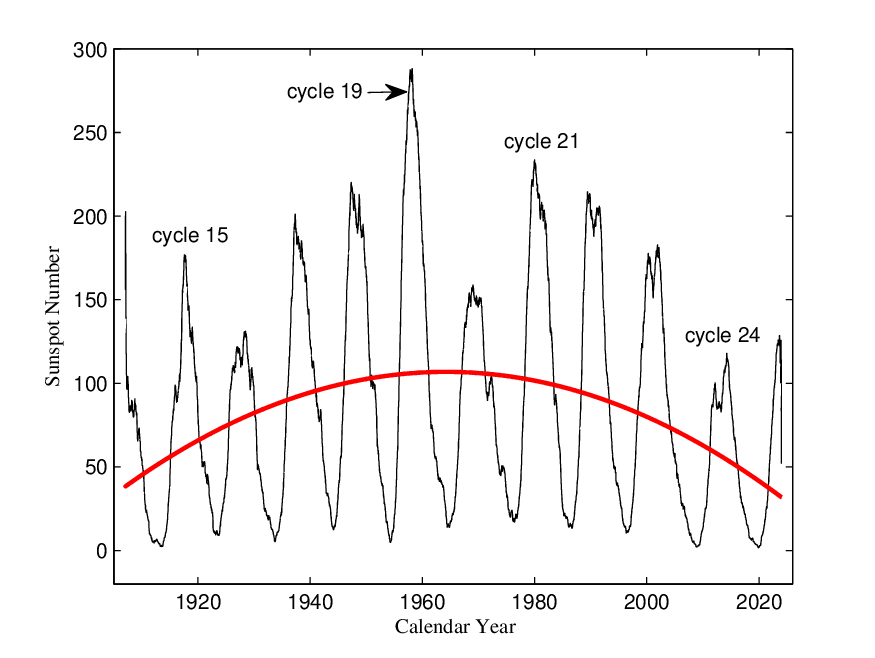}
\includegraphics[width=0.8\textwidth]{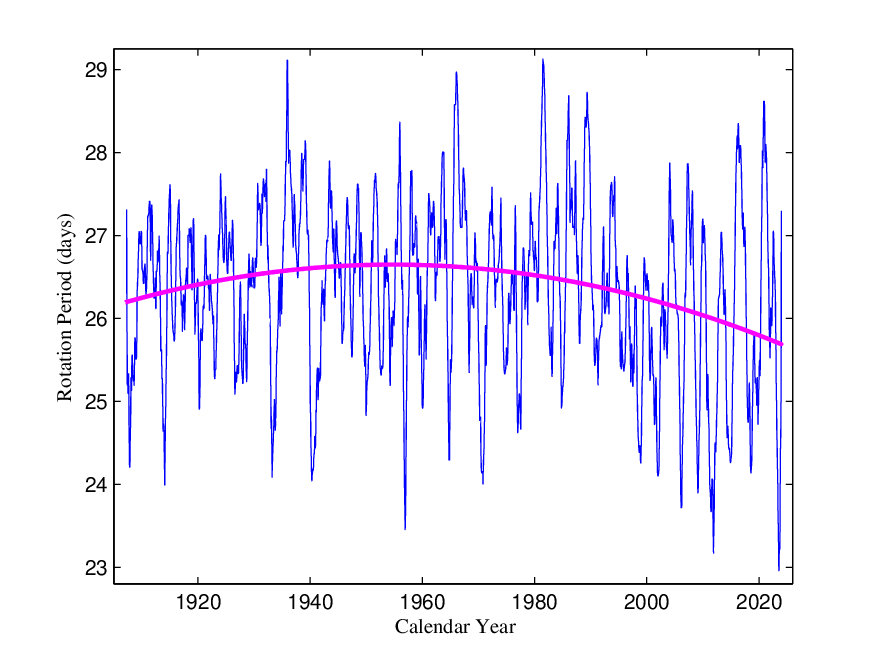}
\end{center}
\caption{Top panel: the 1-year smoothed sunspot numbers from 1907 February 1 to 2023 December 31  (solid black line), and the quadratic polynomial fit (solid red line). Bottom panel: temporal variation of rotation period of the plage area from 1907 February 1 to 2023 December 31 (solid blue line), and the quadratic polynomial fit (solid magenta line). }
\end{figure*}

Prior studies on the long-term trends of solar RPLs using various solar activity indicators have predominantly employed simple linear regression analyses, potentially because the analyzed time series span only a few decades (Chandra \& Vats 2011; Xu et al. 2020; Zhang et al. 2023), and solar activity has been progressively weakening during the recent solar cycles 21-24. However, the dataset we analyzed spans 117 years, covering the descending phase of solar cycle 14 through the ascending phase of solar cycle 25, during which the intensity variations between solar cycles were complex. Therefore, we first conducted a detailed analysis of the variations in solar activity intensity, and then examined the long-term trend of chromospheric rotation based on these results. The daily sunspot numbers (SN) from 1907 to 2023, obtained from the Solar Influences Data Analysis Center, were processed using an 1-year sliding window, with the results depicted by the solid black line in the top panel of Figure 3. As shown in the figure, the SN, an indicator of the level of solar magnetic activity, exhibits complex variations from 1907 to 2023. Nevertheless, the overall trend can be summarized as follows: solar cycles 15-19 generally show a strengthening of solar activity (despite the relatively weak cycle 16), while cycles 21-24 display a gradual weakening. Based on this scenario, we used a quadratic polynomial fit to broadly characterize the overall trend of SN, with the fit result represented by the solid red line in the upper panel of Figure 3. The fitted quadratic polynomial can be expressed as
\begin{equation}
Y_{SN}=-0.021*t^{2} + 82.38*t - 8.08\times10^{4},
\end{equation}
where $Y_{SN}$ represents the fitted SN and $t$ denotes the time in years. This fit was performed using 42,703 data points. An F-test assessing the overall fit of the regression model indicated that it was statistically significant, $F(2, 42701) = 1132.47, p < 0.001$. The statistical test results demonstrate that the fit can significantly represent the overall trend of SN.

Based on the analysis of the SN trend, we performed a quadratic polynomial fit to the RPLs. In the local wavelet power spectra (Figure 2), the values at the beginning and end may not accurately reflect the chromospheric rotation period due to boundary effects. Therefore, during the fitting process, we excluded 200 data points from each end, resulting in a total of 42,303 data points used for the analysis. The fit is shown as the solid magenta line in the bottom panel of Figure 3 and is given by
\begin{equation}
Y_{period}=-0.0002*t^{2} + 0.7819*t - 737.56,
\end{equation}
where $Y_{period}$ is the fitted rotation period in days, and $t$ is the time in years, which corresponds to the variable $t$ in Equation (1). The F-test confirms that the overall fit of the regression model is statistically significant ($F(2, 42301) = 1132.47, p < 0.005$),  demonstrating that the quadratic fit significantly represents the overall trend of the RPLs. As shown in Figure 3, the two fitted curves representing the SN and the RPLs exhibit highly consistent variation trends, with a correlation coefficient of 0.861 indicating a  high correlation. Thus, the long-term trend of chromospheric rotation generally follows the trend of solar magnetic activity levels. Specifically, from solar cycle 15 to 19, the RPLs show a gradual increase, suggesting a progressive deceleration of chromospheric rotation; whereas from cycle 19 to 24, the gradual decrease in RPLs indicates a progressive acceleration of chromospheric rotation.

\begin{figure*}[h!]
\begin{center}
\includegraphics[width=1.0\textwidth]{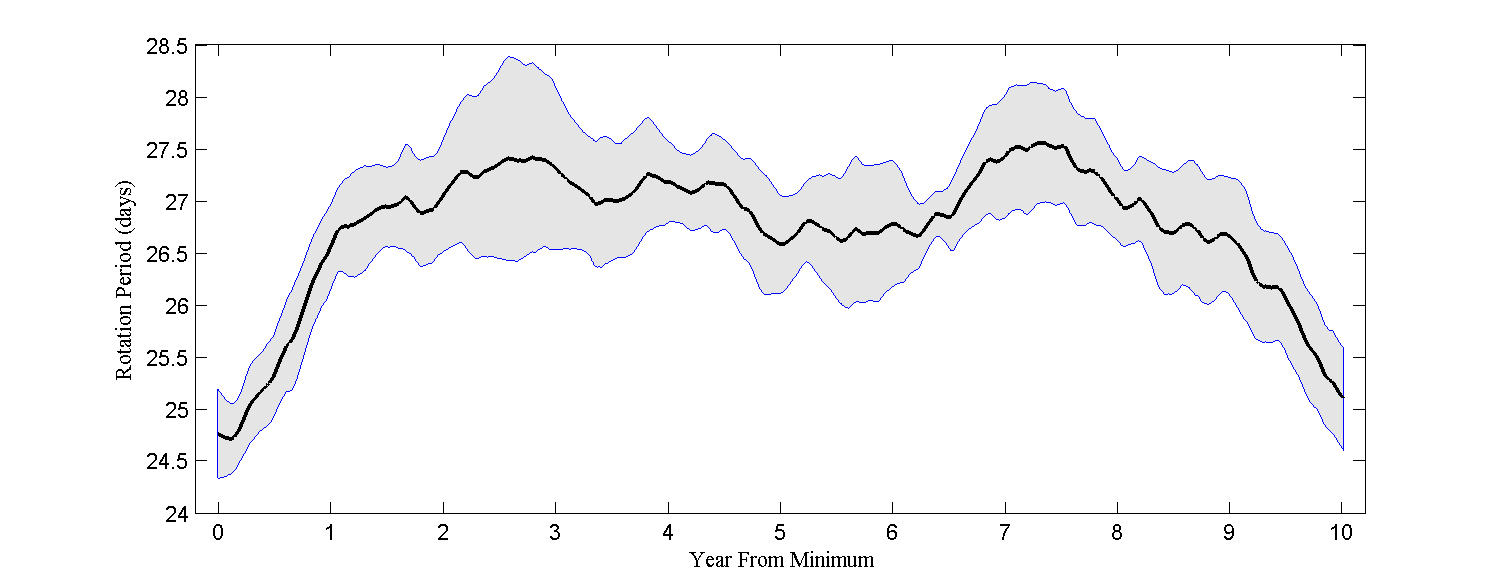}
\includegraphics[width=1.0\textwidth]{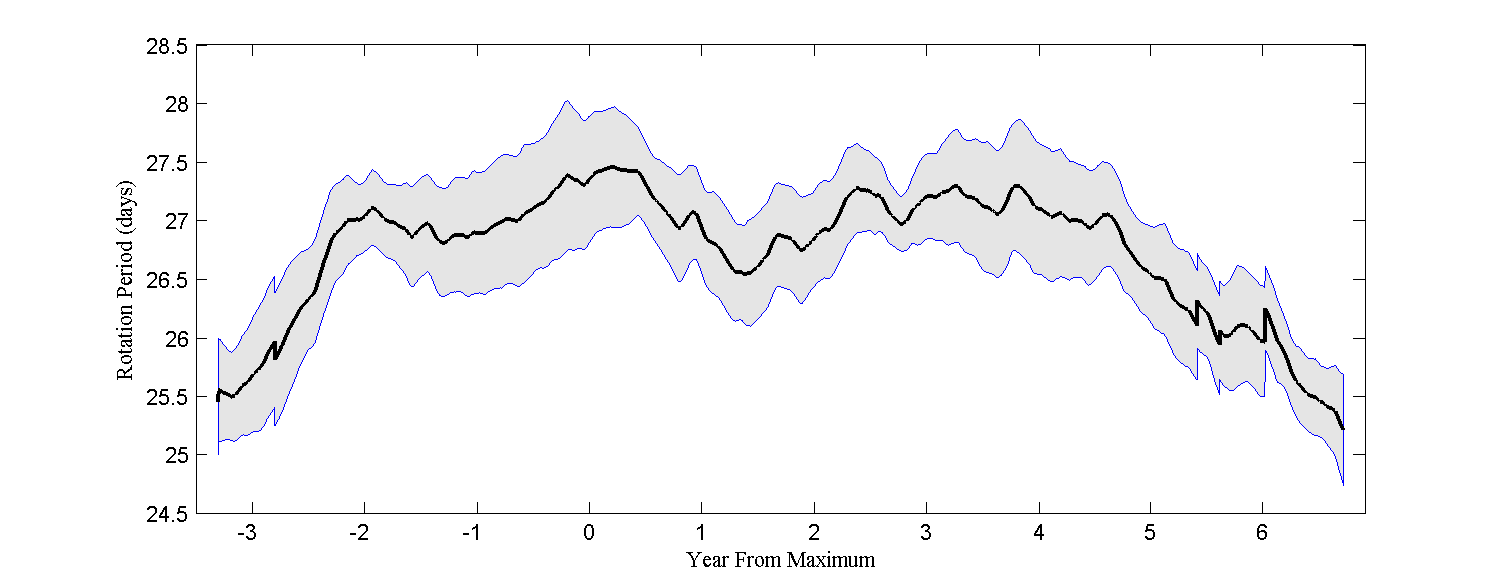}
\end{center}
\caption{Top panel: the dependence of RPLs for plage area on the phase of the solar cycle relative to the nearest preceding sunspot minimum. Bottom panel: same as the top panel, but relative to sunspot maximum. In each panel, the blue lines indicate their corresponding standard errors.}
\end{figure*}

By averaging the RPLs of the time series of the solar activity indicators for the same solar cycle phase (which is defined relative to the nearest preceding sunspot minimum or maximum), we can reveal how the RPLs depends on the solar cycle phase. Thus, based on the results shown in Figure 3 (solid blue line ), the dependence of the  RPLs of the plage area on the solar cycle phase relative to both the nearest preceding sunspot minimum and the sunspot maximum can be obtained, and the results are displayed in Figure 4. As shown in this figure, the RPLs of the plage area is shortest at the initial phase of the solar cycle. It progressively lengthens over the first three years, peaks around solar maximum, and then persists at a relatively long value, exhibiting a double-peak  structure, during the period of relatively strong magnetic field activity  (approximately years 3 to 7.5). Finally, it gradually shortens during the declining phase until it returns to a minimum near the subsequent solar minimum. This points to a negative correlation between the rotation rate of the chromosphere and solar magnetic activity. In other words, the chromosphere exhibits systematically faster rotation during solar minimum phases than during solar maximum phases.

\subsection{ Periodicity in the Temporal Variation of the Rotation Period Lengths of the Chromosphere}
The autocorrelation analysis has been widely used in previous studies to investigate the temporal variation of the RPLs for various solar activity indicators (Chandra \& Vats 2011; Xie et al. 2017; Xiang et al. 2020). Given its established utility, we employ this method to determine  the periods of the RPLs of the plage area. We calculate the autocorrelation coefficient of the RPLs with respect to itself ranging from 1 to 5000 days, which reveals how the autocorrelation varies over this period. To reveal the peaks more clearly, Figure 5 omits the high autocorrelation coefficients at the initial phase shifts, and the y-axis is displayed from -0.15 to 0.8. The two horizontal dashed lines indicate the 99\% confidence level. As shown in Figure 5, the periods of the RPLs for the plage area are approximately 3.2, 5.7, 7.7, 10.3, and 12.3 years, all of which are statistically significant at the 99\% confidence level. The 3.2-year period is likely associated with quasi-biennial oscillations (QBOs). The periods of 5.7 and 7.7 years may correspond to harmonics of the solar activity cycle and the approximately 22-year magnetic activity cycle, respectively. The 10.3- and 12.3-year periods correspond to the approximately 11-year solar cycle.

\begin{figure*}[!htbp]
\begin{center}
\includegraphics[width=0.7\textwidth]{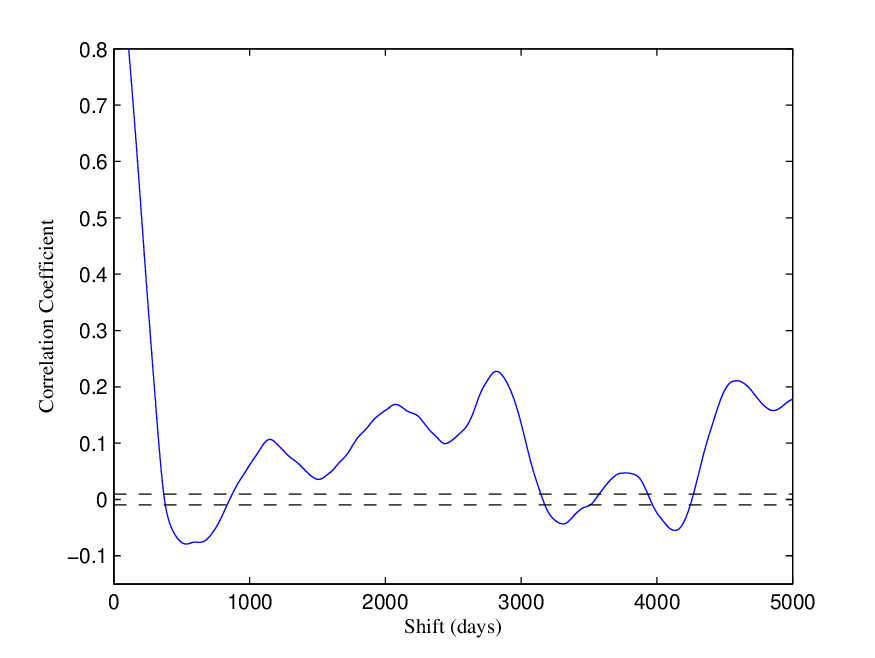}
\end{center}
\caption{Autocorrelation coefficients of the RPLs for relative phase shifts ranging from 1 to 5000 days. The two horizontal dashed lines indicate the 99\% confidence level.}
\end{figure*}

\subsection{Relation of the Rotation Period Lengths of the Chromosphere with Solar Activity}
The SN represents not only the first solar activity data to be systematically observed and recorded, but also the most commonly used indicator for describing the level of solar activity. As such, it is widely employed in solar physics research (Hathaway et al. 1994; Solanki 2003; Hathaway 2015; Xie et al. 2017; Singh et al. 2021a). The time series of daily SN  from 1907 to 2023 was used to investigate the modulation of solar activity on the RPLs of the plage area. The method employed was cross-correlation analysis, which is described in detail in Chandra \& Vats (2011), Xu \& Gao (2016), and Xiang et al. (2020).

\begin{figure*}[!htbp]
\begin{center}
\includegraphics[width=1.0\textwidth]{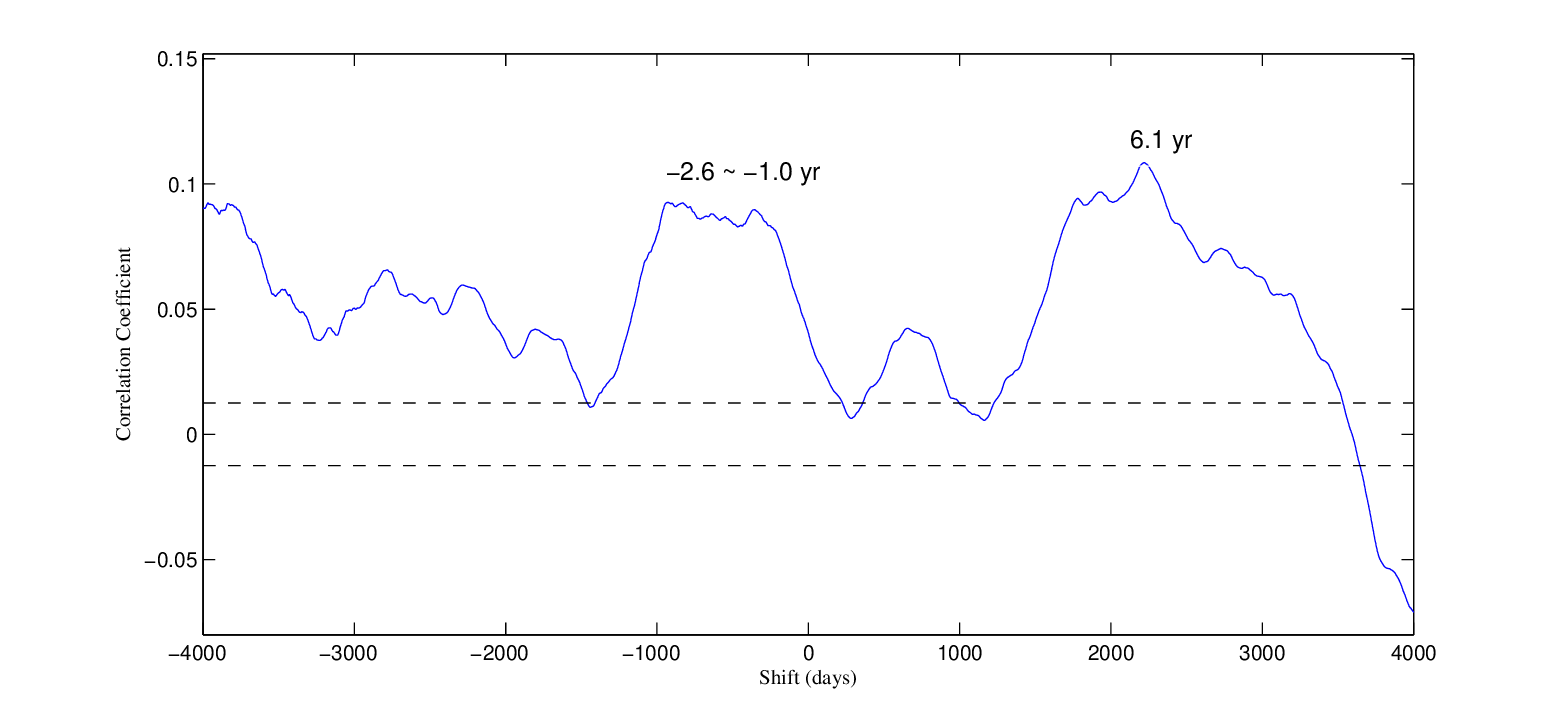}
\end{center}
\caption{Cross-correlation coefficients for the relative  lags between the RPLs and the SN. The two dashed horizontal lines denote the 99\% confidence level, and correlation coefficients lying outside this boundary are considered statistically significant.}
\end{figure*}

The results of the cross-correlation analysis of the relative lag between the RPLs  and the SN are shown in Figure 6. In this figure, the abscissa represents the lag of the RPLs relative to the SN, with negative values indicating a lag of the RPLs. As it shows, the correlation coefficients show multiple statistically significant peaks, suggesting a multi-scale modulation of solar activity on the rotation period of the chromosphere and indicating a highly complex relationship between them. Especially, a maximum correlation coefficient of 0.1085, which is statistically significant, is observed when the RPLs lead the SN by approximately 6.1 years. This peak correlation suggests that the rotation period of the chromosphere generally leads the corresponding solar cycle by about 6.1 years. In contrast to this scenario, the cross-correlation coefficients that are statistically significant exhibits a broad peak in the phase lag range of -2.6 to -1.0 years, where the negative lag indicates that RPLs lags behind the sunspots, accompanied by multiple minor fluctuations within this peak. This reveals that, within this range of phase lags, there exists not only a multi-scale relationship between the chromospheric rotation period and solar activity but also a complex modulation of the rotation period by solar activity.

\section {Discussions and Conclusion}
The  longest available composite time series of the plage areas  from 1907 February 1 to 2023 December 31, that was updated by Chatzistergos et al. (2024), are used to extract the rotation periods using the CWT. Subsequently, we further investigate the multi-scale evolution of chromospheric rotation on centennial timescales and elucidate the physical implications of these evolutionary patterns. In this study, all determined rotation periods are independent of latitudinal differential rotation and are interpreted as global latitude-averaged values; consistent with previous studies, the analysis adopts a global perspective in investigating solar rotation (Heristchi \& Mouradian 2009; Singh et al. 2021a; Xiang et al. 2024). However, The global perspective method is unaffected by factors such as fine-scale structures, localized distributions, and short-term evolutionary features, enabling it to track the continuous changes in the chromospheric rotation period with high temporal resolution.

The continuous wavelet power analysis of the composite plage area time series reveals a temporal evolution of the chromospheric rotation period over the 117-year interval under consideration, with a dominant synodic period of approximately 26.62 days. This period is shorter than the typical synodic period of about 27.4 days for large-scale magnetic activity in the photosphere (Balthasar 2007; Xiang et al. 2014; Li et al. 2019), indicating that the chromosphere rotates slightly faster than the photosphere. This conclusion is consistent with several recent findings (Li et al. 2020, 2023; Wan et al. 2022; Mishra et al. 2024; Rao et al. 2024). In the chromosphere, due to low plasma $\beta$ values and the resulting strong magnetic freezing effect, a magnetically dominated regime exists where plasma motions are constrained by the magnetic field topology (Gary 2001; Wiegelmann et al. 2014). Therefore, the plasma motion in the chromosphere is governed by the combined effect of large-scale magnetic activity regions and ubiquitous small-scale magnetic fields.
In general, the small-scale magnetic concentrations exhibit faster rotation than the large-scale structures , and the rotation rate of the photosphere plasma is slower than that of magnetic structures on the solar surface (Howard 1984; Xiang et al. 2014;  Xu \& Gao 2016). The chromosphere thus rotates faster than the underlying photospheric plasma and large-scale magnetic structures, which is attributed to the effect of small-scale magnetic concentrations (Li et al. 2020; Rao et al. 2024).

Over the past century, the long-term trend of chromospheric rotation is well-characterized by a statistically significant quadratic polynomial, showing a gradual deceleration from solar cycles 15 to 19, followed by a gradual acceleration from cycles 19 to 24. In contrast to previous studies that employed simple linear fittings, our analysis, conducted over a considerably longer time interval, more accurately reveals the long-term evolutionary trend of chromospheric rotation. The accelerated rotational trend observed in recent decades is consistent with findings from other solar indicators, such as SN, Mg II index, Ly $\alpha$ emission, and 10.7 cm radio flux; the majority of these studies attribute this shared phenomenon to the progressive weakening of the sun's global magnetic activity (Heristchi \& Mouradian 2009; Chandra \& Vats 2011; Li et al. 2012; Xie et al. 2017; Deng et al. 2020; Xu et al. 2020; Zhang et al. 2023). However, the gradual deceleration of chromospheric rotation from solar cycles 15 to 19 was first identified in this study. This newly discovered phenomenon is also closely linked to the level of magnetic activity. In fact, the relationship between the solar rotation rate and the level of magnetic activity has been extensively studied in recent decades. Data from sunspots, sunspot groups, and large-scale global magnetic fields, obtained across different periods, consistently reveal that the solar atmosphere rotates faster during periods of low magnetic activity than during periods of high activity (Aini Kambry \& Nishikawa 1990; Hathaway \& Wilson 1990; Obridko \& Shelting 2001, 2016; Braj{\v{s}}a et al. 2006). It has long been observed that large sunspot groups rotate more slowly than their smaller counterparts, and that strong magnetic fields generally inhibit surface rotation (Howard 1984; Xiang et al. 2014; Xu \& Gao 2016). When more large-scale magnetic structures (sunspots) emerge from beneath the photosphere, the rotation of the entire solar atmosphere decelerates, and vice versa. Governed by the strong magnetic freezing effect, chromospheric rotation is modulated by solar magnetic activity. Therefore, its long-term trend must be consistent with the aforementioned relationship between solar rotation and the level of magnetic activity. This expectation is fully confirmed by our results, which in turn provide further confirmation of this relationship from the perspective of the chromospheric rotation.

As evidenced by the dependence of the  RPLs of the plage area  on the solar cycle phase, these RPLs exhibit a negative correlation between the rotation rate of the chromosphere and solar magnetic activity. It is known that the small-scale magnetic concentrations rotate fast than the sunspots (Howard 1984; Xu \& Gao 2016), and the average rotation rate of sunspots is generally higher during solar activity minima than during periods of high activity (Gilman and Howard 1984; Khutsishvili et al. 2002; Braj{\v{s}}a et al. 2006). During solar minima, with very few sunspots present on the solar surface, the variation of chromospheric rotation is mainly modulated by the faster-rotating small-scale magnetic concentrations, resulting in low RPL values, which correspond to faster rotation. The same result has also been found in early studies: research using other solar activity indicators, such as sunspots and the global large-scale field, likewise indicates that the solar atmosphere rotates faster during solar minima (Hathaway \& Wilson 1990; Obridko \& Shelting 2001; Braj{\v{s}}a et al. 2006). During solar maximum, approximately years 3 to 7.5 as shown in Figure 4, more sunspots frequently emerge on the solar surface. The modulation of chromospheric rotation is predominantly governed by the relatively slower-rotating, large-scale magnetic structures. This leads to higher RPL values, which indicate slower rotation. In detail, the gradual and slight decrease in the RPLs during solar maximum corresponds to the migration of sunspots toward lower latitudes. Furthermore, in some solar cycles, the SN exhibits a double-peak structure, which leads to the double-peak structure in the RPLs. In essence, the physical interpretation of this phase-dependent behavior of the RPLs lies in the relative dominance among multi-scale magnetic concentrations with different rotation rates in modulating the variation of chromospheric rotation.

We have also investigated the periodicity in the temporal evolution of the RPLs and have identified statistically significant periods of 3.2, 5.7, 7.7, 10.3, and 12.3 years. The 0.6-4 year periods, considered as QBOs in solar magnetic activity and various  activity indicators, are believed to derive from the base of the solar convection zone, with their effects extending from the solar interior to the heliosphere (Kane 2005; Vecchio et al. 2012; Bazilevskaya et al. 2014; Xiang et al. 2020). The 3.2-year period of the RPLs is thus considered as QBOs. Previous studies on the periodicity of rotational evolution in different solar activity indicators have reported similar periods: for instance, 3.1- and 3.9-year periods in photospheric sunspot data (Javaraiah \& Komm 1999; Li et al. 2011a), 3.0-year in the coronal 10.7 cm radio flux (Xie et al. 2017), and 3.25-year in the modified coronal index (Deng et al. 2020), respectively. The results of this study further confirm that the QBOs are present in the temporal evolution of chromospheric RPLs. That is to say, from the photosphere to the corona, the temporal evolution of their RPLs  shows the continuous  existence of QBOs, which are all modulated by the QBOs of the underlying photospheric magnetic activity. The 5.7-year period detected in the RPLs is the half-subharmonic of the approximately 11-year solar cycle, while the 7.7-year period likely corresponds to the 1/3 subharmonic of the 22-year solar magnetic reversal cycle (Hale's cycle). Furthermore, the approximately 11-year period in the variation of solar rotation rate has been detected in solar activity indicators from both the photosphere and the corona (Javaraiah \& Komm 1999; Obridko \& Shelting 2001; Xie et al. 2017; Deng et al. 2020; Singh et al. 2021a; Sharma et al. 2024). Therefore, the 10.3-year period we discovered in the temporal variation of chromospheric rotation should be related to the 11-year solar cycle. In addition, it is well known that the solar cycle deviates from a strict 11-year period. The length of a given cycle often exceeds or falls short of 11 years; for instance, solar cycle 20 lasted for 11.6 years (Braj{\v{s}}a et al. 2006; Hathaway 2015). Therefore, the 12.3-year period observed in the RPLs may be associated with  the relatively longer solar activity cycle, within the margin of error.

The cross-correlation analysis reveals a complex phase relationship between the RPLs and the solar cycle, indicating that variations in chromospheric rotation are modulated by the solar cycle, and their relationship is highly intricate. The most significant phase difference is that the chromospheric rotation period leads the solar cycle by approximately 6.1 years. The similar leading phase relationship is observed in coronal rotation. Based on the 10.7 cm radio flux, Xie et al. (2017) found that coronal rotation leads solar activity by 5 years. Meanwhile, in a highly intricate manner, the coronal rotational component leads the solar activity component by 5 years, as reported by Li et al (2012). Furthermore, it was also found that the RPLs lag behind the solar activity with a phase lag of 1 to 2.6 years. As shown in Figure 6, the broad range of lags (the wide peak) likely corresponds to the QBOs of solar activity. This implies both a multi-scale complex relationship between the solar magnetic QBOs and variations in chromospheric rotation, and a multi-scale modulation of the latter by the former. Such complex phase relationship  can likely be attributed to  the intricate dependence of chromospheric RPLs on the phase of the solar activity cycle, as discussed  in the aforementioned results. It will be interesting to further investigate this complex relationship in the future.

\begin{acknowledgements}
The authors acknowledge with gratitude Chatzistergos et al. for the creation and dissemination of the composite time series of Ca II K plage area data. This work is supported by the National Natural Science Foundation of China (Grant Nos. 12373059, 12203054, 12463009, 42474224), the Innovation Team Project of Yunnan Revitalization Talent Support Program (202405AS350012), the Yunnan Fundamental Research Projects (Grant No. 202301AV070007), the Sichuan Science and Technology Program (2023NSFSC1349), the Beijing Natural Science Foundation (1242035), the Project Supported by the Specialized Research Fund for State Key Laboratories, and the Chinese Academy of Sciences.
\end{acknowledgements}

\end{document}